\documentclass[fleqn,12pt,twoside]{article}
\usepackage[headings]{espcrc1}

\readRCS
$Id: espcrc1.tex,v 1.2 2004/02/24 11:22:11 spepping Exp $
\usepackage{graphicx}
\usepackage{epsfig} 
\usepackage[figuresright]{rotating}

\newcommand{\AmS}{{\protect\the\textfont2
  A\kern-.1667em\lower.5ex\hbox{M}\kern-.125emS}}
\def\simge{\mathrel{%
   \rlap{\raise 0.511ex \hbox{$>$}}{\lower 0.511ex \hbox{$\sim$}}}}
\def\simle{\mathrel{
   \rlap{\raise 0.511ex \hbox{$<$}}{\lower 0.511ex \hbox{$\sim$}}}}

\title{Recent results on saturation and CGC}

\author{K. Itakura\address{
Institute of Particle and Nuclear Studies, 
High Energy Accelerator Research Organization (KEK), 1-1 Oho, Tsukuba, 
Ibaraki 305-0801, JAPAN.
}%
        }

\runtitle{Recent results on saturation and CGC}
\runauthor{K. Itakura}

\begin{document}

\maketitle

\begin{abstract}
I discuss recent results on the Color Glass Condensate which is
a dense saturated gluonic state and appears as the universal 
picture of hadrons or nuclei at very high energies.
\end{abstract}
\vspace*{-1mm}
\section{Introduction}
\vspace*{-2mm}
Since the previous Quark Matter conference held in January 2004, 
we have seen a very rapid progress in understanding the physics 
of the Color Glass Condensate (CGC).
There are mainly three reasons for this sudden progress:  
First, because new experimental results reported by the BRAHMS 
collaboration in QM04 strongly suggested the existence of the CGC. 
Good experiments always stimulate theorists and lead to 
theoretical developments.
Second, because a new interpretation of the Balitsky-Kovchegov (BK) equation 
was proposed based on the analogy with reaction-diffusion 
dynamics. This gave us an intuitive understanding 
about the emergence of the saturation scale and the geometric scaling. 
Third, because it has been recognized that the BK equation is 
not complete in that it does not contain the effects of 
pomeron loops. This fact drived people to think hard about the physics 
beyond the BK equation. These three reasons are all triggers for the recent 
rapid progress. Indeed all of such activities happened within one 
or two years and some of them are still going on now. Many papers were 
produced during this short period. Thus, instead of covering 
all these activities, I will have to focus on only a few subjects 
in this talk. But, in order to make this talk as self-contained and 
comprehensive as possible, I will start with presenting our motivation 
why we study the CGC. Then I will explain the properties of our basic 
equation, the BK equation, in an intuitive way based on the analogies 
with population dynamics and reaction-diffusion dynamics. 

\subsection{Color Glass Condensate as the high energy limit of QCD} 
The most fundamental and general question which motivated our 
activities is "{\it What is the high-energy limit of QCD?}."
Here "high-energy limit" is meant for the limit of large scattering energy. 
As I explain below, there are enough reasons to believe that 
there exists qualitatively
different picture for hadrons or nuclei when the scattering energy is 
asymptotically large. If it indeed exists, then the natural questions 
to be asked next by experimentalists and theorists
may be, respectively, "{\it Is it already seen in experiments at
current energy?}", and 
"{\it How can we treat it? Can we use weak-coupling techniques?}".
These are the questions which we always have in our minds, and 
I will give answers to them in this talk. But, before doing so, 
let me first introduce two important experimental results which 
suggest the possible form of the high energy limit of QCD. 

The first one is $F_2(x,Q^2)$ structure function measured 
in deep inelastic scattering (DIS) at HERA \cite{New_HERA}. 
The data show steep rise of $F_2$ with decreasing 
$x$, which is ordinarily attributed to the 
increase of gluon density in a proton. Since going smaller $x$ 
corresponds to increasing scattering energy, this implies that
high density gluons can be seen in a highly accelerated proton.  
The next experimental result is the hadronic cross section at high energy.
For example, total cross section for proton-proton scattering, that
is available in the particle data book \cite{PDG} 
for quite a wide range of energy,
shows very slow growth with increasing energy. 
The most recent PDG followed the analyses by the COMPETE collaboration
\cite{COMPETE} and 
adopted the parametrization
$\sigma^{ab}=Z^{ab}+B\ln^2(s/s_0)+\cdots$ whose second term gives 
the leading energy dependence of the data. Here, $\sigma^{ab}$ is 
the (total) cross section for scattering between hadrons $a$ and $b$, 
$Z^{ab}$ is just a constant,
and $s$ is the total energy squared. This form of the cross section 
is motivated by the Froissart bound which is a result of unitarity 
in the $S$-matrix theory. Also important is the fact 
that the coefficient $B$ is universal, that is, independent of 
the species of scattering hadrons. 

These two expermental results suggest that the limit of large scattering
energy will be significantly different from the ordinary picture of 
hadrons, and will be characterizd by {\it many gluons}, {\it unitarity}, 
and {\it universal picture}. In a frame where most of the total momentum is 
carried by the target, the target can be treated as a state having 
these three properties. Recently, this new state of matter which 
becomes relevant at high energy has been named as the Color Glass 
Condensate \cite{review}. 
This name is after the following observations. 
First of all, it is made of gluons ("small-$x$ gluons")
which have {\bf color} and carry small fractions $x\ll 1$ of the 
total momentum of the hadron.
Next, these small-$x$ gluons are created by slowly moving color 
sources (partons with larger $x$) which are distributed randomly 
on the  two dimensional disk (the Lorentz contracted hadron). 
This situation is very similar to that of a {\bf glass} whose 
constituents are disordered and appear to be frozen in short time scales. 
Lastly, the density of small-$x$ gluons becomes very large until 
it is saturated to some value. Typically the occupation number of gluons 
is of ${\cal O}(1/\alpha_s) \gg 1$ at saturation, 
which is like a {\bf condensate} of bosons. 
As the scattering energy is increased, the hadrons undergo
multiple production of small-$x$ gluons, and eventually become
the CGC. Therefore, one can say that {\it the high energy limit of QCD 
is the Color Glass Condensate}.
Note that this claim is as correct as the statement
about the other limits of QCD: high 
temperature/density limit of QCD is the QGP/color superconductor,
which now everyone believes true. In the same sense, if one goes 
to high energy limit in QCD, one will necessarily encounter the CGC. 
Note also that these three different limits allow for
weak-coupling descriptions powered by sophisticated 
resummation schemes.

\vspace*{-3mm}
\section{The Balitsky-Kovchegov equation}
\vspace*{-2mm}
Properties of the CGC are specified by correlation functions of gluons
(or Wilson lines made of gluon fields), 
and change of the CGC with increasing energy is 
determined by "evolution" equations for these correlation functions. 
In particular, the 2-point correlation function determines 
basic properties of the CGC and the evolution for this 
is given by a nonlinear integro-differential equation called 
the {\it Balitsky-Kovchegov (BK) equation} \cite{BK}. 
Physically, the 2-point correlation function corresponds to the 
{\it scattering amplitude} ${\cal N}$ of a "color dipole" off the CGC, 
and can be identified with the gluon number in the target (CGC) when the gluon
field is not strong.
Therefore, the BK equation describes the change of gluon number in a 
target under the change of scattering energy. 
 Explicitly, it is given by ($\bar\alpha_s={\alpha_sN_c}/{\pi}$)
\vspace*{-1mm}
\begin{eqnarray}
\frac{\partial}{\partial Y}{\cal N}_Y(x_\perp,y_\perp)
&=&\frac{\bar\alpha_s}{2\pi}\int d^2z_\perp 
\frac{(x_\perp-y_\perp)^2}{(x_\perp-z_\perp)^2(y_\perp-z_\perp)^2}
\\
&\times&\!\!\!\!\left\{
-{\cal N}_Y(x_\perp,y_\perp)+
{\cal N}_Y(x_\perp,z_\perp)+{\cal N}_Y(z_\perp,y_\perp)
-{\cal N}_Y(x_\perp,z_\perp){\cal N}_Y(z_\perp,y_\perp)
\right\},\nonumber
\end{eqnarray}
where  
${\cal N}_Y(x_\perp,y_\perp)$ is the scattering amplitude 
of the $q\bar q$ dipole with $x_\perp$ and $y_\perp$ being the 
transverse positions of the quark and the antiquark, and
$Y\sim \ln s$ is the rapidity.

As a result of extensive investigation  
of this equation both in analytic \cite{Levin_Tuchin,IIM,MT} 
and numerical \cite{BK_numerical} methods,  
it turned out that there exists saturation regime whose borderline 
is given by the {\it saturation momentum} $Q_s(x,A)$ so that gluons 
having transverse momenta lower than $Q_s(x,A)$ are saturated. 
Intuitively, it corresponds to (inverse of) the typical transverse 
size of gluons when the transverse plane of the target
 is filled with gluons. 
Since the saturation momentum depends upon energy (or $x$) and 
number of nucleons $A$ as $Q_s^2(x,A)\propto A^{1/3}(1/x)^\lambda$ 
with $\lambda\simeq 0.3$ \cite{Dionysis}, 
it grows with increasing energy ($x\to 0$) or for 
large nuclei, and the kinematical region for saturation expands. 
This particular dependence upon $x$ and $A$ leads to an interesting 
observation that the saturation scales for the {\it ep} DIS at HERA 
and for the Au-Au collisions at RHIC are of 
the same order $Q_s(x\sim 10^{-4},A=1)\simeq Q_s(x\sim 10^{-2},A\sim200)$. 
Therefore, if one finds saturation effects in the HERA data, then there 
is enough reason to expect similar things in the RHIC data.
Since most of the gluons have their transverse 
momenta around $Q_s(x,A)$, the weak-coupling treatment becomes 
better and better with increasing energy $\alpha_s(Q_s)\ll 1$. 
Also, the solution at saturation regime (at large rapidities) 
is robust against the small perturbation of the initial condition
specified at lower rapidities. Thus, the saturation regime appears 
to show a universal behavior.
Lastly, the solutions to the BK equation exhibit
new scaling phenomena called {\it geometric scaling} \cite{Levin_Tuchin} 
which 
naturally comes out due to the presence of a saturation momentum, 
and is also observed in experimental data in a beautiful way \cite{GS}.
These properties can be intuitively understood by the analogy with 
population dynamics and reaction-diffusion dynamics, as I explain 
below.

\subsection{Global energy dependence -- the population dynamics}
In order to qualitatively understand what happens in the BK equation, 
let us ignore the transverse dynamics for the time being.
This simplification allows us to find an interesting analogy 
with the problem of population growth \cite{ICHEP}. 
Long time ago, Malthus discussed  
that growth rate of population should be proportional to 
the population itself, and proposed a simple linear equation 
for the population density $N(t)$:
\begin{equation}
{dN(t)}/{dt}=\alpha N(t).
\end{equation}
Its solution $N(t)\!=\!N_0\, {\rm e}^{\alpha t}$ shows exponential 
growth known as the "population explosion." Of course everyone knows 
that such an abnormal future is not our own.
Indeed, as the number of people increases, this 
equation fails to describe the actual growth because  
various effects such as lack of foods help to reduce the 
speed of growth. One can simulate such effects by 
replacing the growth constant $\alpha$ by $\alpha(1-N)$ which 
decreases with increasing $N$. This yields the famous 
{\it logistic equation} which was first proposed by Verhulst:
\begin{equation}
{dN(t)}/{dt}=\alpha \left(N(t)-N^2(t)\right).
\end{equation}
Compared to eq.~(1), this equation has a nonlinear term with a 
minus sign. In Fig.~1, we show the solutions to eq.~(3)
with different initial conditions at $t=0$, together with 
the corresponding solutions to the linear equation (2).
We can learn much from this simple result. First of all, at early time 
$t \ll 1/\alpha$,
the solution shows rapid exponential growth as in the linear case.
However, as $N(t)$ grows, the nonlinear term ($\sim N^2$)
becomes equally important, and the speed of growth is 
reduced. Eventually at late time, the solution approaches to 
a constant ({\it saturate!}) which is determined by the asymptotic condition 
$dN/dt=0$. Next, note that two solutions of the logistic equation 
with different initial conditions approach to each other, and 
converge to the same value, while deviation of two solutions of 
the linear equation expands as time goes. Namely, in the 
logistic equation, the initial 
condition dependence disappears as $t\to \infty$. In other words, 
the solutions to the logistic equation show {\it universal behavior} 
at late time.
\begin{figure}[ht] 
\vspace{-0.4cm} 
\hspace*{6mm}
\parbox{0.57\textwidth}{ 
\epsfig{figure=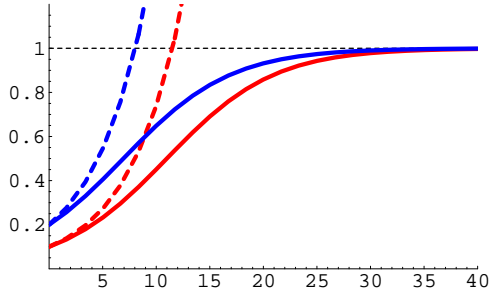,height=0.25\textwidth}} 
\hspace*{-18mm}
\parbox{0.455\textwidth}{
\caption{\small{Comparison of the solutions $N(t)$ to 
eq.~(3) [solid lines] and eq.~(2) [dashed lines], 
with different initial conditions at $t=0$. The figure shows 
saturation and initial-condition independence of the logistic 
solution at late time.}} } 
\vspace{-0.4cm}
\end{figure} 

The similarity of these equations to our problem is rather 
trivial: $N$ and $t$ correspond to the scattering amplitude ${\cal N}$ 
and the rapidity $Y$, respectively. 
When the gluon density is not so high, change of the 
gluon number (scattering amplitude) with increasing rapidity 
is described by the BFKL equation \cite{BFKL}:
\begin{equation}
\frac{\partial}{\partial Y}{\cal N}_Y(k_\perp) 
= \bar\alpha_s K_{\rm BFKL}\otimes {\cal N}_Y(k_\perp),
\end{equation}
where  
${\cal N}_\tau(k)$ is the momentum representation of 
the dipole scattering amplitude, and 
$\bar\alpha_s K_{\rm BFKL}$ is the kernel 
representing the probability of splitting of one dipole into two.
This is essentially a linear equation, and its 
solution at asymptotically large $Y$ shows exponential 
growth ${\cal N}_Y(k)\sim \exp\{(4\bar\alpha_s \ln 2)\, Y\}$. 
This result is an analog of the population explosion in the 
population dunamics.
This solution, however, violates the unitarity bound for the amplitude 
(${\cal N}_\tau\le 1$) and the BFKL equation must be 
modified so as not to violate the unitarity. In fact, what is missing 
in the BFKL equation is the {\it recombination process} of two gluons 
into one, which cannot be ignored 
when the gluon density is high. Note that this process 
effectively reduces the speed of growth. 
Once this is included, the BFKL equation is replaced by 
the BK equation (1). 
In the momentum space, it can be schematically represented as
\begin{equation}
\frac{\partial}{\partial Y}{\cal N}_Y(k_\perp)=
\bar\alpha_s K_{\rm BFKL}\otimes \left(
{\cal N}_Y(k_\perp)-{\cal N}^2_Y(k_\perp)\right)\!.\!\!
\end{equation}
Notice the similarity in the structure with the logistic equation (3).
Therefore, it is now easy to understand that the solution to the BK 
equation shows (i) {\it saturation} and {\it unitarization} 
of the gluon number and (ii) {\it universality} that the solution at 
very large rapidity becomes independent of the initial condition.

\subsection{Traveling waves in the reaction-diffusion dynamics}
The intuitive picture presented above is not just an analogy, 
but can be justified as the limiting case of the remarkable observation 
made by Munier and Peschanski \cite{Munier}. In a series of papers, 
they established the following fact:
\begin{itemize}
\item[] \hspace*{-8mm}{\bf Fact}:
{\it Within a reasonable approximation, the BK equation in the momentum space 
(5) can be rewritten as the FKPP (Fisher, Kolmogorov, Petrovsky, Piscounov) 
equation}\vspace*{-2mm}
\begin{equation}
\partial_t u = \partial_x^2 u + u-u^2
\end{equation}\vspace*{-2mm}
{\it where} $t\sim Y,\ x\sim \ln k_\perp^2$ {\it and} 
$u(t,x)\sim {\cal N}_Y(k_\perp).$  
\end{itemize}
The FKPP equation is a famous equation in non-equilibrium statistical 
physics covering many interesting phenomena such as directed percolation, 
pattern formulation, spreading of epidemics, etc., and has been 
investigated in great detail. The dynamics described by this equation 
is called the {\it reaction-diffusion dynamics} because the last two 
terms represent "reaction" part which is equivalent to the right-hand 
side of the logistic equation (3) with $\alpha=1$, while the first 
term represents the diffusion. (Now it is easy to understand that 
the logistic equation (3)
indeed comes out under the constant mode approximation 
$\partial u/\partial x =0$.) 
Therefore, the solution to this equation is determined due to the 
interplay between these two effects. As we saw before, the logistic 
part induces the transition from unstable (exponentially growing) state 
to stable (saturated) state at some position $x$. 
On the other hand, the stable region expands due to the effect of 
diffusion. Therefore, it is very natural 
that the FKPP equation has a {\it traveling wave} solution. Typical 
traveling wave solutions at different time $t$ and $t'>t$ are shown 
in Fig.~2. 
\begin{figure}[ht] 
\vspace{-3.5cm} \hspace{4mm}
\parbox{0.55\textwidth}{ 
\epsfig{figure=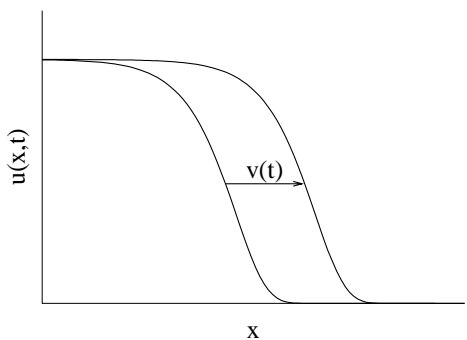,height=0.6\textwidth}} 
\hspace{-14mm}
\parbox{0.49\textwidth}{
\caption{\small{Traveling wave solutions to the FKPP equation (6)
at different time $t$ (left)
and $t'>t$ (right) as a function of $x$. (Figure by courtesy of 
R.~Enberg)}} } 
\vspace{-3.5cm}
\end{figure} 

There are two facts about this traveling wave solution,
both of which are significantly relevant to the saturation physics:
\vspace{-1mm}
\begin{itemize}
\item[] \hspace*{-8mm}{\bf Fact I} : {\it For a traveling wave solution, 
one can define the position of a "wave front" $x(t)=v(t)t$, irrespective of 
the details of the nonlinear effects.}\vspace{-1mm}
\item[] \hspace*{-8mm}{\bf Fact II}: 
{\it At late time, the "shape" of a traveling wave is preserved 
during its propagation, and the solution becomes only a function of 
$x-vt$.}
\end{itemize}
\vspace{-1mm}
Let us translate these facts into the language of saturation physics.
Fact I is one of the most important facts for the saturation physics 
because {\it the position of the wave front is nothing but the saturation 
scale} $x(t)\sim \ln Q_s^2(Y)$.
By definition, the position of the wave front locates at the transition 
point from the unstable to stable regimes. On the other hand, the 
saturation scale is defined as the boundary between saturated (stable) and 
non-saturated (unstable) regimes. Therefore it is quite natural that 
the position of the wave front corresponds to the saturation scale. 
What is more important is that the saturation scale can be determined 
irrespective of the nonlinear effects. This is again consistent with 
our knowledge that the energy dependence of the saturation scale can 
be determined by the linear BFKL evolution. 

Fact II itself implies that the solution $u(t,x)$ which is originally 
a function of position $x$ and time $t$ will show a scaling that it 
depends only on a specific combination of two variables $x-x(t)$. 
Within the saturation language, one finds that this phenomenon 
corresponds to the {\it geometric scaling} where the scattering amplitude 
becomes a function of a particular combination of transverse momentum 
$k_\perp$ and some function of $x$, namely, the scaling variable is 
$x-x(t)\sim \ln^2 k_\perp^2/Q_s^2(Y)$ and thus 
${\cal N}_Y(k)=f(k_\perp^2/Q_s^2(Y))$. This scaling holds very well 
in a deeply saturated regime, where the profile of the solution does 
not change. This again implies the universality of the saturation 
regime. On the other hand, as one departs from the saturation regime 
(or the wave front) toward dilute (unstable) regime, 
the effect of saturation becomes weaker and weaker and eventually 
disappears, and the solution 
ceases to show the scaling. Still, one can approximately see the 
scaling if one stays close to (but outside of) the saturation boundary.
One can estimate the upper limit of the transverse momentum squared 
below which the solution will show the scaling. Namely, the scaling
is approximately seen in the following window (refered to as the 
extended scaling regime) \cite{IIM}:\vspace{-1mm}
\begin{equation}
Q_s^2(x)\ \simle \ Q^2\  \simle\ Q_s^4(x)/\Lambda^2_{\rm QCD}.
\end{equation}
\vspace{-1mm}This upper limit is roughly consistent with the experimental data at 
HERA \cite{GS}.
The geometric scaling is indeed observed up to $Q^2\sim 100 {\rm GeV}^2$ while
the saturation scale at HERA is estimated as about $Q_s^2\sim 1 {\rm GeV}^2$
at $x\sim 10^{-4}$.

Hence we recognized that there is a qualitatively different regime 
in between the CGC and dilute regimes. The theoretical status at this 
point for a proton as seen in DIS is summarized in Fig.~3. 
\begin{figure}[ht] 
\vspace{-1.9cm} \hspace*{-6mm}
\parbox{0.5\textwidth}{ 
\epsfig{figure=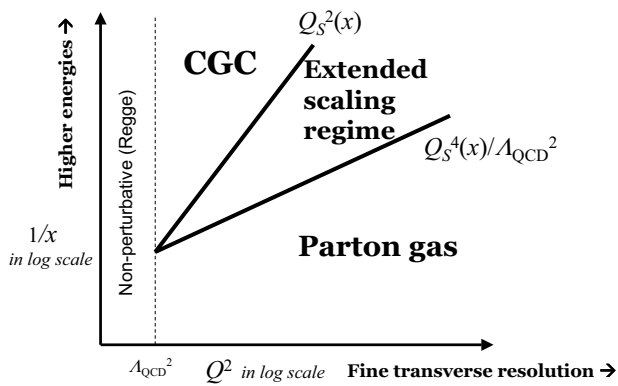,height=0.45\textwidth}
} 
\hspace{18mm}
\parbox{0.33\textwidth}{
\caption{\small{"Phase diagram" of a proton as seen in 
deep inelastic scattering.}} 
} 
\vspace{-1.9cm}
\end{figure} 

\section{Recent progress in phenomenology}
Our understanding of the CGC has been deepened partly (in fact, largely) 
due to the experimental results at HERA and RHIC. Below I briefly explain 
some of the attempts to describe/understand the experimental data 
from the viewpoint of the CGC. \vspace{-1mm}
\subsection{DIS at HERA}
DIS at small $x$ is the cleanest process for measuring 
saturation effects in gluon distribution of the proton.
Starting from the pioneering work by Golec-Biernat and 
W\"usthoff~\cite{GBW}, there are several attempts to describe the 
HERA DIS data \cite{New_HERA} 
in the context of gluon saturation. 
Here let me briefly explain the "CGC fit" \cite{CGC_fit} 
which is one of the most 
successful fits of the small $x$ data based on QCD.
The CGC fit is constructed so as to contain two approximate 
solutions to the BK equation, which are valid 
in the saturation and BFKL regimes, respectively.
In particular, the solution in the linear BFKL regime 
shows the geometric scaling and its (small) violation. 
With only three parameters, the CGC fit provides a very nice 
fit for the $F_2$ structure function with $x<10^{-2}$ and 
in $0.045 < Q^2 < 45\,$GeV$^2$. From the fit, the saturation scale 
was determined as \vspace{-1mm}
\begin{equation}
Q_s^2(x)=(1\, {\rm GeV})^2 (x_0/x)^{\lambda},\  \ {\rm with}\  \ 
x_0=0.26\times 10^{-4},\ \lambda=0.25.
\end{equation}\vspace{-1mm}
Meanwhile, it turned out that
this fit works reasonably well even for other observables such as 
$F_2^{\rm diff},\ F_{\rm L}$ and the vector meson production 
\cite{Forshaw,Victor}. Precise determination of the saturation scale
 is also important for the analysis involving nuclei because  one can
determine the saturation scale for the nucleus by using the result
for a proton.

\vspace{-1mm}
\subsection{Au-Au collisions at RHIC}

The CGC provides the initial condition for the heavy ion collision. 
Information of the initial state could  still be seen in the 
final observed data. It should be noticed that most of the produced 
particles have small momenta less than 1 GeV which is of the same order 
as $Q_s$ in RHIC. This observation suggests that 
effects of saturation may be visible in bulk quantities such as 
the multiplicity. Indeed, the CGC results \cite{Dima_Levin,Hirano_Nara} 
 for the pseudo-rapidity and centrality dependences of the multiplicity
are in good agreement with the data.
\vspace{-1mm}
\subsection{Deuteron-Au collisions at RHIC}

Going to forward rapidities in a p-A collision corresponds 
to probing a nuclear wavefunction at smaller $x$, which should 
exhibit saturation with decreasing $x$. Thus, this is one of 
the best places to search for the CGC or the effects of quantum 
evolution \cite{Forward}. For example, such effects should be 
measured in the nuclear modification factor, and this was 
indeed done 
by the BRAHMS experiment in the deuteron-Au collisions at RHIC \cite{Brahms}. 
The experimental data show {\it enhancement} of the ratio at 
mid-rapidity (the Cronin effect) and {\it suppression} at 
forward rapidities. Such global behavior is qualitatively 
consistent with the predictions made by the CGC \cite{Dima,Armesto}.
After the data was announced, many publications followed
to confirm that this phenomena are indeed due to the saturation 
and CGC (for a review, see Ref.~\cite{Kovchegov_Jamal}). 
For instance, detailed analytical investigation 
of the ratio\footnote{More precisely, a ratio of the nuclear 
wavefunction to the proton wavefunction scaled up by $A^{1/3}$, 
which shows very similar behaviors as the nuclear modification factor.} 
was performed by Iancu, Triantafyllopoulos, and myself \cite{IIT} 
and it has been clarified that the 
Cronin effect is due to multiple Glauber-Mueller scattering 
and re-distribution of gluons, both of which are properly 
described by the McLerran-Venugopalan model (classical saturation 
model without evolution), 
and that the high $p_\perp$ suppression is induced by the 
mismatch of the evolution speed between the proton (deuteron) 
and the nucleus. The nucleus is closer to saturation and thus evolves 
slower than the proton. Quantitative results are also available.
Kharzeev, Kovchegov, and Tuchin have computed the nuclear modification 
factor within the framework of the CGC \cite{KKT}, 
and found rather good agreement 
with the BRAHMS data. Very recently, there was an important progress:
Dumitru, Hayashigaki, and Jalilian-Marian recognized that including 
the DGLAP evolution in the {\it projectile} side significantly improves 
the transverse spectra of the produced particles \cite{DHJ}. Up to now, their 
result seems to be the best one for this observable. Also important 
is the fact that the averaged value of $x_A$ (the fraction of gluon 
momentum coming from the target nucleus) is small enough 
$\langle x_A\rangle \sim 10^{-3}$ for the $2\to 1$ kinematics, 
which is highly contrasted with the results 
$\langle x_A\rangle \sim 10^{-2}$ for the standard 
$2\to 2$ kinematics with the leading twist shadowing 
\cite{2to2}. Therefore, our framework gives 
a consistent description of the deuteron-Au scattering.

Lastly, various observables have been computed and found to show 
suppression due to saturation. They include dileptons and photon 
productions \cite{EMprobes} for the electro-magnetic probes, 
$q\bar q$ or 
heavy meson productions \cite{meson}, 
jet azimuthal correlations \cite{jet-jet}, etc.
Note that EM probes are important in that they are less ambiguous 
because the process does not contain fragmentation functions. 

So far, we have understood many things within the framework of 
the CGC, but in fact there are several other approaches which are aimed 
at describing the Cronin effect and high $p_t$ suppression. 
Thus, in order to be convinced enough, it is necessary to perform 
more detailed investigation in the future.

After these phenomenological analyses, we can add numbers on the 
axes of "phase diagram" in Fig.~3. The results are summarized
in Fig.~4, where the diagrams for a proton and a nucleus 
are shown separately because respective saturation scales 
 are different. Since the saturation scale (squared) for Au is 
larger than that for a proton by a factor $A^{1/3}\sim 6$,
the saturation regime for Au is bigger.   
Kinematically allowed regions for HERA and RHIC are also 
specified on the figure. 
Clearly, HERA and forward rapidities at RHIC have large overlapping 
with saturation regime, while the mid-rapidity at RHIC does not.

\begin{figure}[ht] 
\vspace{-1.7cm} \hspace*{-16mm}
\parbox{0.47\textwidth}{ 
\epsfig{figure=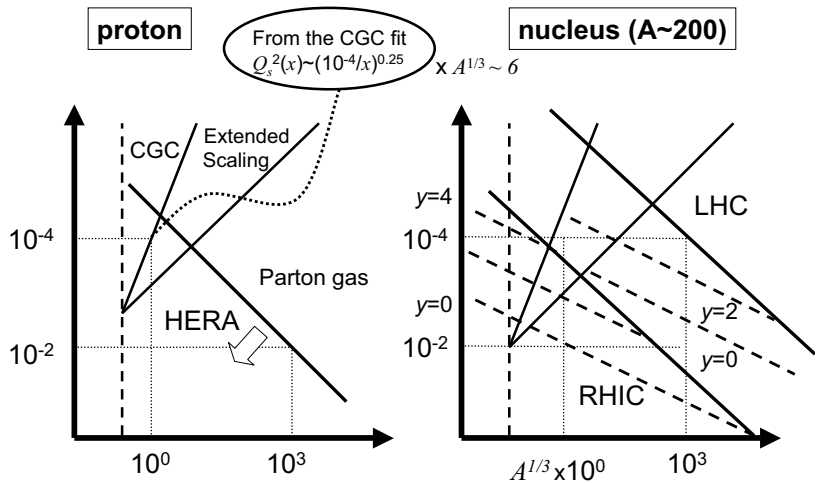,height=0.56\textwidth}
} 
\hspace{46mm}
\parbox{0.33\textwidth}{
\caption{\small{"Phase diagram" with numbers determined from the 
phenomenological analyses. Vertical and horizontal axes are 
$x$ and $Q^2$ (or $k_\perp^2$ of gluons 
for a nucleus) in log scale. Kinematically allowed regions are
shown for HERA, RHIC and LHC experiments.}} 
} 
\vspace{-2.3cm}
\end{figure}

\subsection{CGC at the LHC}

Now it is straightforward to draw lines for the LHC
on the same phase diagram (number of nucleons for Pb 
is not so different from that for Au). As is seen in 
Fig.~4 (right), kinematically allowed 
region for the LHC has significant overlapping with the saturation 
regime even at mid-rapidity. More precisely, for the same transverse 
momentum, the saturation scale at the LHC is increased by a factor 3 
than that of RHIC. Therefore, the effects of saturation is expected 
to be more visible at the LHC. There are already predictions 
for the multiplicities for pA and AA collisions \cite{LHC} and 
the nuclear modification factor for pA collision
\cite{KKT}. Predictions for other observables are also
necessary before the LHC starts to operate. In order to make 
realistic predictions, we will have to consider both the 
initial-state (CGC) and final-state (energy loss) effects.
Even at the RHIC energy, the similar situation should be seen 
at forward rapidities in Au-Au collisions. Studying this situation 
in RHIC will be very helpful to understand the future LHC experiments.

\vspace{-2mm}
\section{Recent progress in theory -- Beyond the BK equation}
\vspace{-1mm}
Most recently, there is a growing acceptance that the BK equation 
is not sufficient to correctly describe the high-energy limit 
of QCD. Since this was first discussed in detail by Mueller and Shoshi
\cite{beyondBK}, studies of the physics beyond the BK equation is 
becoming one of the main subjects of the CGC or saturation physics. 
Research on this subject is still rapidly developping with some 
(technical and conceptual) problems left unresolved, and 
it is rather difficult (and even not appropriate) at this time 
to make a conclusive statement (for references, see the citation
list of Ref.~\cite{beyondBK}: there are 51 hits by now). 
Instead, I would comment briefly 
on some general picture which we are aiming at.

Suppose that we have a complete description of the high-energy 
limit of QCD. It should at least contain 
pomerons (2 reggeized gluon exchange, C even), 
odderons (3 reggeized gluon exchange, C odd), and 
reggeons (quark antiquark exchange, etc) as the exchanges between 
a projectile and a target, 
and correct interaction among them. 
On the other hand, the BK equation describes only (multiple exchanges of) 
pomerons, and pomeron {\it merging} as the interaction 
(from the target point of view).
This implies that the BK equation is {\it not symmetric} under the exchange
between a target and a projectile. In order to obtain a symmetric picture
which also contains other exchanges, we have to go beyond the BK equation.
This activity has two different aspects: one is to consider 
$n$-point correlation functions ($n>2$), and the other is to find
correct interactions among the above-mentioned exchanges.
Note that the evolution of the CGC can be formulated as a stochastic 
process governed by a Hamiltonian. Within this context, the pomeron 
 can be described as a kind of two particle collective state 
(a two point function) of 
this Hamiltonian. Recently, it has been established that the 
odderon exchange also appears as the collective state of this 
Hamiltonian, but is given by a "three particle" state 
(a three point function)
which is odd under charge conjugation \cite{Odderon}. However, 
this Hamiltonian itself has to be modified because it does not 
contain the pomeron {\it splitting}. Inclusion of pomeron splitting 
to the evolution equation has been discussed by several people. 
What I can say at present is that we are certainly approaching 
towards the complete description of the high-energy limit of QCD.

\end{document}